# Thermal Management in Large Data Centers: Security Threats and Mitigation

Betty Saridou[1], Gueltoum Bendiab[2], Stavros N. Shiaeles[2], and Basil K. Papadopoulos[1]

[1]Democritus University of Thrace, Xanthi, Greece
{dsaridou, papadob}@civil.duth.gr
[2]Cyber Security Research Group, University of Portsmouth, Portsmouth, UK
gueltoum.bendiab@port.ac.uk, sshiaeles@ieee.org

***Abstract.*** **Data centres are experiencing significant growth in their scale, especially, with the ever-increasing demand for cloud and IoT ser- vices. However, this rapid growth has raised numerous security issues and vulnerabilities; new types of strategic cyber-attacks are aimed at specific physical components of data centres that keep them operating. Attacks against temperature monitoring and cooling systems of data centres, also known as thermal attacks, can cause a complete meltdown and are generally considered difficult to address. In this paper, we focus on this issue by analysing the potential security threats to these systems and their impact on the overall data center safety and performance. We also present current thermal anomaly detection methods and their limitations. Finally, we propose a hybrid method that uses multi-variant anomaly detection to prevent thermal attacks, as well as a fuzzy-based health factor to enhance data center thermal awareness and security**.

## 1 Introduction

Data Centres are experiencing unprecedented growth and will continue to scale operations to meet the ever-increasing service demands. A recent study by Gartner[1] estimated that about 425 million new servers might be needed by 2020 to support 30 billion connected IoT devices around the world, while the average data centre will support more than 100,000 servers [30]. The average data warehouse size usually ranges from 100ft$^2$ to 400,000ft$^2$ [35] and operates on a high- power consumption. A plethora of security issues has arisen due to this rapid growth, causing data centres to become vulnerable to strategic cyber-attacks on the physical infrastructure vital to maintaining the uninterrupted operation of data centres [12, 13, 19, 39]. Power supply, cooling, temperature monitoring and even security systems can serve as entry points for attacks against data centre operators or companies using data centre services. Security experts warn that ne sophisticated malware such as Triton and Trisis [28] are very effective against power and HVAC (Heating, Ventilating, and Air Conditioning) systems, and therefore, put data centres' safety at risk. Moreover, reported cyber-incidents showed that cyber-attacks on such connected facilities can destroy thousands of servers by overheating the environmental atmosphere, or manipulating energy and temperature settings to cause a fire or an explosion incident [12, 13, 39]. Attacks against temperature monitoring and cooling systems, also known as thermal attacks, are considered dangerous and difficult to tackle.

In this paper, we aim to address this issue by exploring the most important security threats against thermal management systems, as well as their overall impact on data centre security and performance. In this context, we describe a number of vulnerabilities that may be exploited by attackers who want to affect environmental conditions inside a data centre. Based on a combination of thermal hardware requirements, thermal anomaly detection is considered a relevant approach for detecting abnormal behaviour of data centre indoor temperature. In this context, several approaches have been proposed for temperature measurement analysis and abnormal behaviour definition of various environments such as smart homes, health, and data centres. In this paper, we analyse the effectiveness and limitations of these approaches with respect to the protection against thermal attacks in data centres. Additionally, we denote the importance of intelligent solutions and propose a hybrid framework that uses multi-variant anomaly detection as a preventive measure. Finally, to further enhance thermal awareness and security of data centres, we introduce a fuzzy-based health factor.

---

[1] https://www.gartner.com/en

The remainder of this paper is organised as follows. In section 2, a high-level overview of temperature monitoring in large data centres is presented, along with their importance and benefits. Section 3 discusses the potential security threats on temperature monitoring systems and their impact on the overall data centre security and performance. In Section 4, we provide an overview of the existing solutions and their limitations. Section 5 outlines the effectiveness of combining multi-variate anomaly detection methods and fuzzy logic in protecting data centres against thermal attacks. Finally, section 6 concludes this paper and outlines future work.

## 2 Thermal management in data centres

Data centres are large infrastructures composed of a building or a group of buildings dedicated to housing computer systems and their services [13]. They usually host a large number of high-density servers, that are run simultaneously and generate an extensive amount of heat every second. Thus, existing data centres are equipped with multiple cooling technologies to cool down servers, including chilled water systems, cold aisle/hot aisle design, Computer Room Air Conditioner (CRAC), liquid cooling and free air cooling [12, 32]. For efficient thermal management, they also use robust thermal design and temperature regulation and monitoring as heat countermeasures. In modern data centres, HVAC and CRAC systems are the most widely used cooling solution for managing temperature, airflow and humidity, while continuous monitoring with Wireless thermal sensor networks (WSNs) helps to create and maintain the energy-efficient cooling environment.

As shown in Figure 1, computer rooms of the data centre are arranged into hot and cold aisles; server rack fronts face each other to create cold aisles due to the front-to-back heat produced by servers [32]. CRAC units, which are usually positioned at the end of hot-aisles or around the room perimeter (see Figure 1) push the cold air under the raised floor and through the cold aisle to absorb the heat produced by servers. The resulting hot air enters the hot aisle from the front side of the server racks and is returned to the CRAC devices to cool down by the chillers once more [12, 13, 32]. In order to avoid overheating, the inlet temperatures of servers are continuously monitored using a network of thermal sensors. If the inlet temperature exceeds a predefined threshold, parts of servers or even racks are forced to shut down to avoid permanent hardware damage [12].

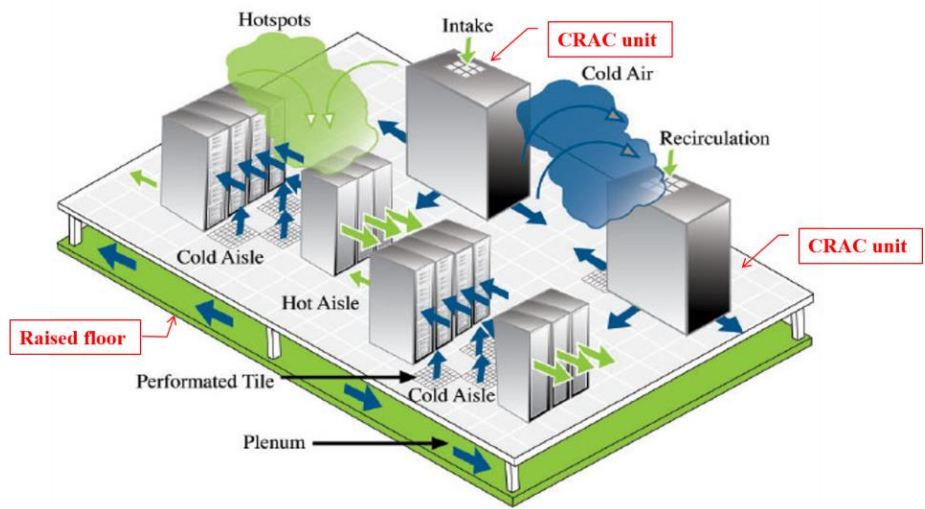

**Fig.1.** Typical cooling system in data centres [11].

In general, control and monitoring of heat diffusion inside a data center is a complicated process that involves rack positioning design, air flow simulations and monitoring devices distribution. The American Society of Heating, Refrigerating and Air-Conditioning Engineers (ASHRAE) supplies detailed suggestions for temperature readings and defines three models regarding placing temperature sensors: a) managing the space, b) setting up the space, and c) troubleshooting the space. More specifically, it suggests that temperature should be measured at every fourth rack position in the center of cold aisles. Additionally, temperature measurement should be taken at the center of the air intakes of the top, middle, and bottom equipment mounted in a particular rack [33]. Temperature monitoring is not only effective against equipment protection, but it is also an efficient measure to cut costs. In 2013, a study by the U.S. General Services Administration proposed a temperature of 22°C – 26°C as the ideal temperature range for data centres. The same study claimed savings of about 4-5% in energy costs from every 1°C of temperature increase [36]. Overall, temperature monitoring provides numerous advantages to data centre managers, offering energy use optimisation, optimal heat distribution, and thermal overload avoidance among others.

## 3   Security threats and their impact

Data centres play an increasingly important role in modern society and digital economy, however; their security is still a challenging area with many open problems. The increasing reliance on these digital infrastructures to accommodate large volumes of valuable information make them prone to strategic attacks. In this context, many studies showed that most data centres and their associated building automation systems have little or no protection policies in place [8, 12, 13, 19]. Further, a recent report by CyrusOne[2] confirmed that its New York data centre was attacked by the REvil ransomware, also known as Sodinokibi, affecting more than six customer companies of CyrusOne [20]. Another report by Verizon found that DDoS attacks and privilege misuse were the most common attack vectors in last year's cyber-attacks against data centres, mainly due to the proliferation of poorly secured connected IoT devices [21].

---

[2] https://cyrusone.com/

The temperature monitoring system, which is one of the most critical components of a data centre, is vulnerable to cyber-attacks too [12, 19]. Temperature monitoring is usually guaranteed through hundreds or even thousands of temperature sensors (i.e., Thermal sensors) that form a WSN, which can then be remotely controlled and configured. Thus, malicious cyber actors can easily take control of the network and manipulate the thermal conditions of a specific server or rack; for example, maintaining servers in a relatively hot environment [12, 13]. To achieve their goals, attackers will run thermal-intensive workloads into victim servers or VMs (Virtual Machines) to rapidly generate large amounts of heat. This heat will consequently worsen the thermal condition of the peripheral environment, and therefore raise the inlet temperature of other servers [12, 14]. What's more, such networks are vulnerable to a wide variety of attacks due to several restrictions, such as low capability of computation, small memory size, limited resources of energy and unreliable communication channels [31]. Overall, these insecure devices can be easily compromised in order to remotely launch attacks against the temperature monitoring system.

The adversary who launches a thermal attack can be an individual hacker, a competing provider, or a specialised organisation for committing cyber-crime. A study by [34] showed that thermal attacks can also come from insider malicious tenants, especially in the case of multi-tenant data centres. In this case, malicious tenants can inject additional thermal loads, exceeding the shared cooling system capacity, which can lead to overheating and possible system downtime. It is clear that the impact of thermal attacks is not limited to victim servers, but it can also affect surrounding environmental temperature, which, in turn, can impact thermal conditions of other servers and the entire data centre [14]. In the same context, a study by [12] found that a successful thermal attack targeting 2% of servers in a data centre can dramatically affect the thermal conditions of the whole data centre. Similarly, affecting thermal conditions of adjacent servers and causing local hotspots would significantly raise cooling costs, or lead to a cooling breakdown. Taking these studies into consideration, it has been well-proved that thermal attacks can highly decrease performance and reliability of victim servers by largely increasing their temperature.

One of the main reasons that renders data centres vulnerable to thermal attacks is the extensive use of aggressive cooling and power management policies such as power oversubscription [12, 39]. Power oversubscription allows more servers to be hosted on a single data centre's power infrastructure than it can actually support [12]. Implementations of power oversubscription can lead to severe security gaps, e.g, malicious attackers can manipulate many servers to create a local hotspot or generate simultaneous power peaks to violate the power capacity in the data centre [39]. Another reason is that thermal attacks try to intentionally keep the outlet temperature of victim servers at a high level, which makes chip-level temperature sensors, used in more modern servers, incapable of monitoring the temperature at the server- and data centre-level. These sensors can only provide information about the server's core temperature which is not equal to the inlet and outlet temperatures, and therefore, cannot prevent the occurrence of local hotspots [12].

Current anomaly detection approaches in data centre temperature monitoring have many weaknesses and are not effective in preventing thermal attacks [12, 25]. For instance, most of these approaches are not capable of distinguishing between benign thermal-intensive workloads and the malicious ones, as several thermal-intensive incidents are benign and do not exploit security vulnerabilities [12]. Furthermore, a study by [13] demonstrated that existing anomaly detection approaches cannot defend against data centre-level thermal attacks, where a large number of accounts run different workloads simultaneously.

# 4 Thermal anomaly detection in data centres

Anomaly detection is a relevant approach in the problem of temperature monitoring inside data centres because it allows data centre managers to identify behaviour that deviates from "normal" system behaviour in a proactive manner. In general, thermal anomaly detection is based on a combination of thermal hardware requirements, that have been applied to cooling systems, and historical observations, that are recorded and/or streamed by environmental sensors inside the critical infrastructure. Several researchers have suggested anomaly detection methods for temperature data in a variety of environments. Initially, and in an attempt to detect possible risk events inside a household, an anomaly detection method was applied to smart home data by [18] through temporal data mining. For their investigation, researchers used sensors to measure temperature, humidity, light, etc. The model looked for temporal interactions among frequent activities to learn "resident" behaviour and reported as "unusual" events with low probability. Despite using a combination of real and synthetic data, the method proved effective, but was not used to perform predictions [18].

Deviations from historical patterns were also examined in environmental data in [15]. The study presented an anomaly detection method, using a data-driven univariate auto-regressive model and a moving window to predict the next measurement based on historical data. The method, which employed sensors reporting through telemetry, performed well without requiring previous classification of data, and proved applicable to larger data sets as well [15]. In their extensive study, [6] proposed a spatiotemporal correlation to detect anomalies from sensor nodes, such as temperature, humidity and light, in wireless sensor networks. Even though this method aimed to reduce energy and spectrum consumption, the information-gain from aggregate neighbourhood sensor data can prove critical in high-risk cases where temperature variations need to be addressed immediately. In an attempt to ensure healthy operating conditions and ensure reliability for High-Performance Computing systems (HPC) and data centres, [5] used autoencoders to train data collected from monitoring devices mounted on computing nodes. Their method used metrics such as core load, power consumption, room temperature, GPU usage, cooling fans speed, etc., and was able to catch anomalies when tested on real tier-1 HPC systems.

Despite the fact that temperature was not one of the monitoring values, re- search conducted by [1, 16] demonstrated the use of anomaly detection in critical infrastructure. On one hand, [16] presented an analysis of multi-variate anomaly detection methods to assist in condition-based maintenance. Their approach included an application on real aircraft turbofan engine data, where their model was able to successfully detect anomalies and recognise failing factors. On the other hand, [1] proposed a multi-agent swarm system to detect abnormal be- haviour in cloud data centres. Nevertheless, both frameworks did not consider data collection from the surrounding environment for their evaluation, but rather focused on the system's parameters instead.

The study of thermal anomaly detection specifically designed for a data centre environment has only started to gain attention during the last decade, where there has been a significant development of data warehouses around the globe. Studies published by Marwah et al. [25, 26, 27] in 2009-10 were the first to ad- dress the issue of autonomous anomaly detection of temperature values in data centres and apply machine learning on sensor data sets. In their first paper on the topic, [27] underlined the importance for an autonomous system that is able to catch temperature values beyond a certain threshold. They also listed common technical reasons, as well as their symptoms, that lead to anomalies. As they state, most of these anomalies can go undetected by traditional systems, while they demonstrate how Principal Component Analysis (PCA) can assist as the primary detection mechanism.

In [25], they extended their initial hypothesis by introducing four detection mechanisms. More specifically, they compared a simple threshold method, a moving average method, an Exponentially Weighed Moving Average (EWMA) method and naive Bayes to predict thermal outliers on a three-month data set. Naive Bayes outperformed the other models and was able to predict 18% of the anomalies at an average of 12 minutes before occurrence. Next, in [26], they used hierarchical PCA for real-time detection, which resulted in a 98% accuracy of predicting anomalous cases. What is more, the method was applied to data coming from a part of the data centre where traditional methods would not be able to raise an alarm.

In [40], researchers introduced a sophisticated two-tier hierarchical neural network framework that detected server-level as well as data centre-level thermal anomalies. The method was able to achieve this by studying the relationships of heterogeneous sensors and consequently outperform other machine learning models. After extensive research on thermal maps for data centres, [22, 23] pro- posed a novel anomaly detection method that compares and maximizes the accuracy of constructed and observed maps to detect Regions of Interest (RoIs). The method, which was notably based on thermal cameras and traditional temperature monitoring devices, demonstrated sufficient accuracy in anomalous cases. Later, and responding to the growing need for fast online detection of fault cases, [3, 41] introduced the use of Self-Organizing Maps and reputation systems.

In another study published in 2016, [4] presented a sophisticated four-step tool based on density estimation for anomaly detection and data exploration, which was specifically designed for HPC sensors. Even though this method was tested and performed well on a real HPC environment, it was not designed for real-time detection, an approach that recent studies tend to follow. [24] focused exclusively on temperature data when they used naive ensembles to detect anomalies in the cooling system of data centres. Nevertheless, because of its theoretical approach, their method needs to be extended to demonstrate sufficient performance on non-simulated scenarios. During the same year, another study published by [9] proposed an architecture sensing scheme. According to it, heat sensors collect server temperature data and transmit them to the cloud for further analysis. Later in 2018, [7] tested several machine learning models on server temperature data to examine cooling reliability of data centres. By constructing workload-independent cooling profiles servers, they were also able to detect both transient and lasting cooling failures of servers. Two extensive studies published by [17, 42] in 2019, deployed machine learning to detect fault operations of HVAC systems in data centres. In [42], researchers recognising the importance of air-cooling systems in data centres introduced a hybrid approach for four- fault decoupling features, including compressor valve leakage, condenser fouling, evaporator airflow reduction, and liquid line restriction. Data were manipulated using the random forest algorithm, which proved rather effective in fault events diagnosis. Lastly and for the same reason, in [17], researchers considered three different types of anomaly detection methodologies, namely naive point anomalies, contextual point anomalies, and level shifts. Machine learning methods were then employed on a real data set showing good precision rates.

Taking everything into consideration, there have been several approaches to analyse temperature measurements and define abnormal behaviour of the respective environments. Recent studies have started to include critical infrastructure in their analyses as well, but their approaches are rather focused on optimal system management, energy efficiency, and reliability of services. Given the recent increase of cyber-attacks in industrial systems, we believe that real-time temperature monitoring in data centres should start to be examined by the security community as an attack mitigation measure. Moreover, temperature monitoring should not be analysed separately, but rather in conjunction with network and security monitoring measures.

Additionally, there is a need for efficient time response of anomaly detection methods in operational processes applied to data centres; this type of approach can be observed in the field of financial streaming analytics, such as in the study of [2], where having the lead in financial transactions provides an advantage. Finally, we should note that the type and quality of data used in model testing is of paramount importance. In the next section, we address these issues by proposing a framework based on a combination of recent methods and tools published by [10, 29, 37, 38].

## 5  Multi-Variant Anomaly Detection & Fuzzy Logic

In an attempt to address the specific technical problem, we propose a hybrid method that merges the strongest elements of the methodologies in [10, 29, 38]. A study by [10] highlighted the need for performing anomaly detection in multi-variate time sensing environments and proposed RADM, a real-time anomaly detection algorithm based on Hierarchical Temporal Memory (HTM) and Bayesian Network (BN). Their methodology outperformed anomaly detection in univariate sensing time-series, when tested on CPU, network, and memory sensors. Additionally, [10] introduced the health factor α, which measures the overall health of the system by considering the individual anomaly scores calculated by HTM for each parameter (e.g., CPU). We should note that HTM is a far more accurate representation of the neural structures and mechanisms of the brain than the widely spread neural networks.

In their study, [29] demonstrated a sophisticated decentralised scheme for fault detection and classification where wireless sensor networks are prevalent. According to their sensor monitoring methodology, neighbour sensors can be grouped by conducting calculations among sensor readings. For sensor measurement, their choice of time-series model was the Auto-Regressive Moving Average model (ARMA). Fault detection is performed when reading is com- pared to a certain threshold and then the fault classification algorithm is initiated to determine the fault type. The fault reading is checked for the frequency and continuity, as well as the presence of an observable pattern. According to these criteria it is then classified as a) random, if discontinuous and appearing randomly, b) malfunction, if discontinuous and appearing frequently, c) bias, if continuous and exhibiting no pattern, d) drift, if continuous and following a pattern. For their investigation, they used an outdoor temperature dataset, achieving accuracy rates between 85% and 95%. The ever-increasing need for intelligent systems in the security domain, also led to non-traditional approaches, such as Fuzzy Logic and Fuzzy Inference Systems. Researchers in [38] proposed a fuzzy-based approach to overcome complexities in Building Energy Management Systems (BEMS). More specifically, they were able to use the nearest neighbour and fuzzy rules to extract normal building behaviour. Then, they used fuzzy linguistic descriptors for a variety of parameters, such as zone temperature, exhaust fan load, supply fan current etc.

In this paper, our goal is to perform anomaly detection with the main focus on temperature while, considering various measured phenomena as inputs. The proposed methodology combines elements from all three previous works and split them into four main steps. First, we perform a grouping of sensors as per the type of measured phenomenon or the spatial distribution of sensors measuring the same reading. Second, we perform fault detection on the grouped time series models individually. In the case where a faulty reading has been detected, we perform fault classification to define its type according to continuity, frequency, and pattern criteria. Third, the grouped time series are inserted into the RADM framework, where the global anomaly region is defined. If once again, the system detects an anomaly on the overall system, we perform fault detection to define the type. Finally, we perform a permutation of the health factor α based on Fuzzy Logic and Fuzzy Rules. More details about each step are provided in the following sections.

## 5.1 Step 1: Thermal Sensors Grouping

In this step, we acknowledge that the ambient temperature of data centres is a product of more than one factors. For instance, new or advanced cooling systems, such as rack cooling, have been employed in recent years to aid in reserving the servers' ideal state. Consequently, there is a clear need to consider more than one input when building time series models for anomaly detection. For this reason, we propose a sensor grouping technique which groups sensor readings that measure the same phenomenon, e.g., grouping of all ambient air temperature sensors in a room, fan motor speed, etc. Similarly, sensors can be also grouped spatially, for example aggregating high-mounted ambient temperature sensors of a specific aisle in a single group, low-mounted ambient temperature sensors in a different one, creating a group for high-mounted rack temperature sensors, and so on. For this method, we employ the Neighbourhood Voting scheme proposed by [29]. According to them, the system does not require a priori knowledge about the of sensor readings. By changing the notation to keep the consistency of our framework, we present the Neighbourhood Voting algorithm as it appears in environment. Instead, it takes advantage of the redundancy in measurements their study:

1) Collect the set of readings $R = r\,[1 \ldots |Neighbour\,(Y_i)|]$ from all neighbours, excluding its own reading $r_i$.

2) Calculate the median of the group, $\mu = \{R\}_{\frac{1}{2}} = \bar{r}$.

3) Calculate the difference between $r_i$ and $\bar{r}$, $D_{r,\bar{r}} = |r_i - \bar{r}|$

4) Compare the difference $D_{r,\bar{r}}$ with a threshold $\tau$, that can be adjusted.
   - If $D_{r,\bar{r}} < \tau$ then $r_i$ is a good reading.
   - If $D_{r,\bar{r}} \geq \tau$ then $r_i$ is a faulty reading. We define faulty readings as $D_{r,\bar{r}} = \varepsilon_i$

where Yi is a node measuring a specific phenomenon, $r_i$ is its reading. The set of its neighbours is denoted by $Neighbour\,(Y_i)$ and the number of neighbours by $|Neighbour\,(Y_i)|$.

## 5.2 Step 2: Fault Detection

In this step, we perform fault detection on the aforementioned grouped measurements. Because of the many advantages it exhibits for anomaly detection over traditional time-series, grouped readings are modelled with HTM time-series as suggested by [10], and not the ARMA model preferred in [29]. If a faulty reading is detected, the Fault Classification Algorithm proposed in [29] is initiated. The algorithm is run parallel to the HMT modelling to classify the faulty measurement as either malfunction, random, bias, or drift, according to the Fault Classification Algorithm (Algorithm 1), where T represents time intervals, $\varepsilon_i$ the numbers of faulty occurrences, and θ the desired threshold.

**Algorithm 1:** Fault Classification [29].

**Input:**
1. $R\,[1..T]$: vector of $T$ sensor readings
2. $E\,[1..T]$ vector of the faulty state of $R\,[1..T]$

**Output:** $C$: fault type of sensor node in the interval

1 compute the occurrences of faults $\varepsilon_i$ in $R$
2 check the continuity
3 **if** $\varepsilon_i$ *is discrete* **then**
4  Check the frequency
5  **if** $|\varepsilon_i| > \theta_i$ **then**
6   $C = Malfunction$
7  **else**
8   $C = Random$
9 **if** $\varepsilon_i$ *is continuous* **then**
10  Check the fault function $\varepsilon_i$
11  **if** $\varepsilon_i = const$ **then**
12   $C = Bias$
13  **else**
14   $C = Drift$
15  return $C$

## 5.3 Step 3: Anomaly Region Detection

During the third step of the process, the HTM time-series of the grouped variables are combined with the Bayesian Network to detect the anomaly region. The time-series follow the RADM framework proposed by [10]. As previously discussed, we define the temperature T of the data centre as a product of factors X, Y, and Z, according to Equation 1:

$$T(t) = (X(t), Y(t), Z(t)) \quad (1)$$

Then, the overall process, including steps 1, 2, and 3, is described in Figure

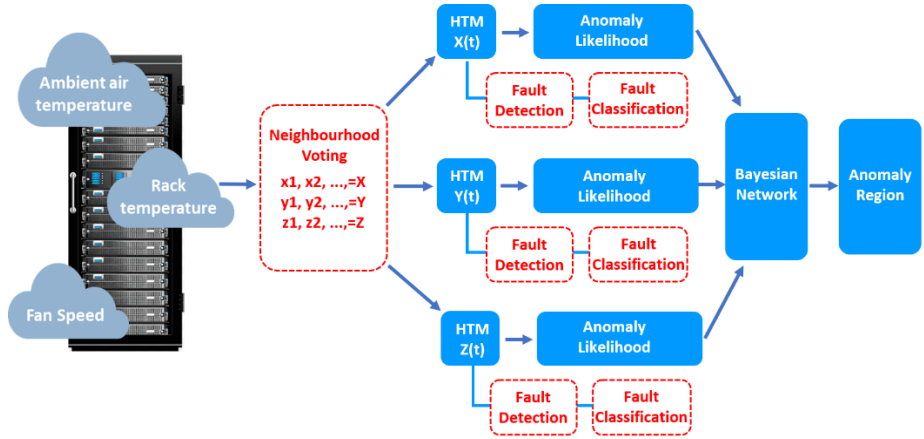

**Fig. 2.** Steps 1, 2, and 3 of the combined methods. Boxes in red represent methods proposed by [29], while boxes in blue comprise the RADM framework proposed by [10].

### 5.4 Step 4: Fuzzy sets

In the fourth step of our methodology, we propose a permutation of the health factor α proposed by [29], in order to assess the overall wellbeing of our frame- work. Our method uses linguistic descriptors to describe the concept of the system's anomaly state in an aggregated and intelligent manner. The method uses fuzzy rules similar to the analysis by [38], except the input variables consist of the anomaly scores Si calculated by the HTM model for the grouped variables. More specifically, the input variables are represented using five fuzzy sets, namely very low, low, medium, high, and very high. The output variable Health is represented again by five fuzzy sets, that describe the state of the system as: *very bad, bad, average, good*, and *very good*. Input and output linguistic descriptors are shown in Figure 3. Fuzzy rules are then constructed according to experience and describe the contribution of individual anomalies to the general state. The linguistic descriptors for the healthiest state with the three inputs used in our example would be:

$$\textbf{IF } S_{X(t)} \textbf{ IS } Very\ Low \textbf{ AND } S_{Y(t)} \textbf{ IS } Very\ Low \textbf{ AND } S_{Z(t)} \textbf{ IS } Very\ Low \textbf{ THEN } Health\ \textbf{IS}\ Very\ Good \quad (2)$$

Finally, to put this methodology into action, we propose the DAD data set published by [37]. The DAD data set is a labelled IoT data set, which contains real-world behaviours of a data centre as seen from the network. The network and environmental values were obtained from a physical data centre, combined with temperature measurements transmitted by NFC smart passive sensor technology.

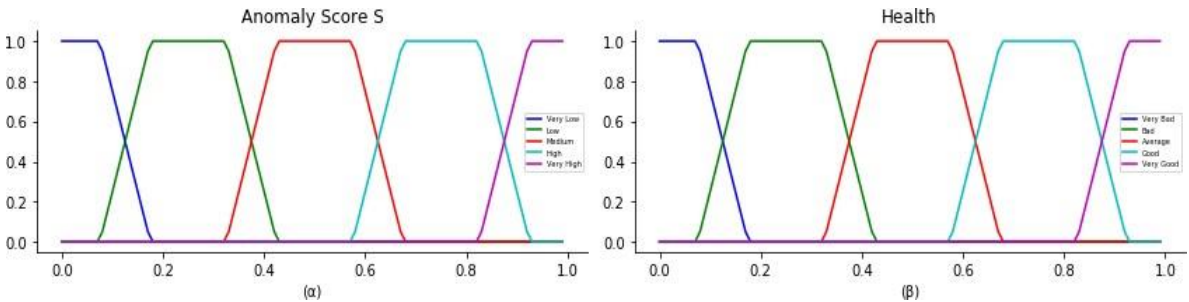

**Fig.3.** Linguistic descriptors for (*α*) individual Anomaly Scores and (*β*) overall Health.

## 6   Conclusions and Future Work

Thermal attacks present a serious vulnerability in data centres today. By launch- ing a thermal attack, an adversary can cause a complete meltdown to the equipment stored inside a data centre. In this paper, we examined how easily temperature monitoring systems can be used by attackers to manipulate heat distribution in such environments. We also presented and reviewed existing model-based anomaly detection methods that have been focused on measuring temperature deviations in critical infrastructures. According to our findings, thermal attacks are fundamentally not addressed as potential cybersecurity attacks for data centres, and consequently, there is a serious lack of frameworks proposed to eliminate this threat. We believe that future studies on the current topic should be focused on constructing intelligent solutions. In our research, we found several approaches to tackle this issue and addressed the most important aspects for a thermal attack mitigation methodology. According to this, we suggested the use of a multi-variate anomaly detection method that could perform fault classification, and a fuzzy-based health factor to assess of the overall state of the system. Finally, we propose a curated data set that could assist in the exploration of the aforementioned models.

## Acknowledgement

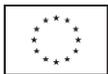 This project has received funding from the European Union's Horizon 2020 research and innovation programme under grant agreement no. 786698. This work reflects authors' view and Agency is not responsible for any use that may be made of the information it contains.